# Transport Enhancement and In Situ Control of Electronic Correlation via Photoinduced Modulation Doping of van der Waals Heterostructures

Collin R. Sanborn*, Son T. Le*, Thuc T. Mai*, Maria F. Munoz, Riccardo Torsi, Angela R. Hight Walker, Curt A. Richter, Samuel W. LaGasse, Aubrey T. Hanbicki, Adam L. Friedman

Modulation doping, a well-established technique for traditional semiconductor heterostructures, is a promising approach for tailoring carrier concentration in 2D materials devices. In this letter we report on photoinduced modulation doping in hBN/graphene/hBN/SiO2 heterostructures utilizing standard white light sources and no additional fabrication complexity. We establish the use of this technique to both dope the channel material and to "photoanneal" devices, providing control over electronic doping and disorder in the graphene channel. We analyze the transport properties by employing Drude and Landauer transport models, highlighting the ability to reversibly tune the mobility and mean scattering length of the graphene with a high degree of accuracy. This tunability allows us to switch our device between the diffusive and quasi-ballistic transport regimes in situ. We utilize the exceptional control our technique provides over local disorder to realize quantum Hall isospin ferromagnetic states in a device whose initial quality would otherwise leave such states unobservable. These results demonstrate precise manipulation of carrier density and charge disorder in van der Waals heterostructures, providing a highly accessible approach to creating high-quality devices capable of realizing correlated electronic states.

Two-dimensional (2D) materials are an exciting platform for next-generation computing devices and condensed matter physics. Studies of the fundamental electronic transport properties of 2D materials rely on precise control over the charge carrier density in active layers such as graphene or 2D semiconductors. To this end, a variety of doping schemes involving chemical adsorbates[1–4], substitutional atoms[2,5], or electrostatic gating[6–9] have been proposed. Chemical and substitutional approaches provide access to high concentrations of dopants, yet require additional fabrication considerations and don't offer practical paths to tunability outside a laboratory environment. Electrostatic gating enables dynamic tuning of the carrier density with a fine degree of control, but requires continuous charging of the gates to function. Additionally, spontaneous trapping of charges can lead to device degradation over time.[6,10] Device-to-device variation of gate threshold voltages is also of major concern for future industrial viability of 2D material-based electronic devices.[11] The strengths of these two doping techniques can be combined in a third approach, modulation doping, whereby dopants that are located remotely from the active layer can provide strong doping without degrading the device performance. We utilize a powerful yet underexplored approach, photoinduced modulation doping, and demonstrate the tremendous potential of the technique to enhance and control the electronic transport properties of nanoscale devices.[12–16]

Modulation doping in traditional device architectures, such as 2D electron gas quantum well devices, has demonstrated tremendous promise towards device enhancement and control.[17,18] Traditional modulation doping approaches in semiconductor heterostructures rely on tailored growth of heterogeneous layers with fine control over dopant density. Modulation doping has been achieved in van der Waals heterostructure systems though integration of novel 2D materials with particular charge transfer characteristics, yielding improved contact resistance and channel doping, but requires precise placement of the 2D layers and adds additional fabrication complexity.[19–23] Here we investigate photoinduced modulation doping achieved by integration of common 2D materials and inexpensive light sources.

Under applied electric field and incident light, the doping level of 2D materials heterostructures consisting of multiple dielectric layers can be precisely controlled and tuned in situ.[12,14] This photoinduced modulation doping – or photodoping – offers local control of charge with submicron spatial extent and lifetimes on the order of months to years.[12,13,24,25] For 2D materials systems, there is no current agreement on the exact mechanism underlying this effect, with studies pointing to charge generation and trapping in various layers/interfaces. Proposals include photoexcited charge generation directly in channel materials, commonly graphene or transition metal dichalcogenides (TMDs)[26–30], in dielectric layers such as hBN and $SiO_2$[12,27], or at interfaces between these materials.[25] Studies of related effects indicate the potential for charge generation within the Si substrate as well.[31–33] Irrespective of the exact nature of charge generation, there is general agreement that free photo-generated electrons and holes may be swept into adjacent layers by the applied field, leading to charging of deep trap states within the heterostructure. Carriers are repeatedly generated and fill trap states until the applied field is fully screened by trapped charges, which remain trapped even after the applied field is removed. Initial studies of the photodoping effect proposed that these traps were within the dielectric bulk[12,15,29], but the majority of recent studies propose trapping at the hBN-$SiO_2$ interface.[24–26,30,34] We note that most studies do not directly probe these generation or trapping mechanisms, and what evidence does exist supports trapping at the dielectric-dielectric interface, with charge generation potentially occurring within each layer depending on the power and spectrum of the illumination used.[27] Additional work suggests this interface trapping is not unique to the most commonly used dielectrics: hBN and $SiO_2$, and systems of other dielectrics merit further research.[25,28,30] Regardless of the mechanistic details, this photodoping effect yields high-quality reconfigurable devices in a variety of 2D material systems including graphene, TMDs, and 2D superconductors.[25,29,35–37]

In this letter, we build upon our previous work showing tunability and determining the spatial extent of the photodoping effect[13]: demonstrating that this technique may be used to enhance electronic transport properties in a model graphene-based device. We perform backgate-dependent transport measurements of an hBN-encapsulated graphene device and demonstrate nonvolatile control over the electrostatic environment via our photodoping approach. Through application of Drude and Landauer transport models to our experimental data, we highlight how this control improves key device parameters such as mobility and elastic scattering length. We then utilize this enhancement to control correlated physics in situ: measuring the quantum Hall isospin ferromagnetic states in graphene with full degeneracy lifting and the ability to toggle the degeneracy on and off via our photodoping technique.[38] Most interestingly, we achieve the exceptional device quality needed to observe these states in a sample whose initial quality would have precluded such an observation. This success underscores the promise of photoinduced modulation doping for achieving precise and tunable control of charge while reducing device variability in 2D materials heterostructure devices.

## RESULTS

To investigate the nature of photodoping in van der Waals heterostructures we constructed an hBN-encapsulated graphene test structure in the Hall bar geometry (Fig. 1a). We choose graphene as our channel material over semiconductors such as TMDs because it has a clearly resolvable peak in the channel resistance at the charge neutrality point (CNP) in a transfer curve measurement (Fig. 1b). Demonstrations of reconfigurable doping in TMDs suggests our results are also applicable to semiconducting channels.[16,20,39] We initially observe transport characteristics of good, but not exceptional, quality. Rather than spending substantial time and resources fabricating another sample, we employed photodoping methods to directly enhance the quality of this one.

The sample was illuminated with white LED light (emission spectrum, Fig. S4) focused on the sample through a 50x objective overnight with the backgate ($V_{\mathrm{BG}}$) and contacts grounded. We define this process as "photoannealing" and note that this process also serves to reset the CNP of the device to its intrinsic value of -3.8 V. To then photodope the sample a fixed voltage, $V_{\mathrm{write}}$, was applied between the backgate and contacts while the heterostructure was illuminated in the same manner for approximately ten minutes. Other broadband light sources, such as halogen lamps, were tested and gave similar results to the LED source. After the light source was extinguished, the backgate voltage was removed and electrical characterization measurements were performed.

To determine the effectiveness of photodoping in our graphene device, we incrementally n-dope the device, taking transfer curve measurements between each photodoping exposure (Fig. 2a). For clarity, when referring to doping in the rest of this work we refer to the induced doping in the graphene channel, and not the photoinduced trapped charges at the hBN-$SiO_2$ interface. We then incrementally step $V_{\mathrm{write}}$ back towards 0 V, discharging the trapped charges and decreasing the n-doping of the channel (Fig. 2b). Our photodoping approach enables precise control over the doping level in the system, as evidenced by the tunability demonstrated in our transfer curve measurements. The charge neutrality point can be repeatedly and reversibly tuned to a specific value over a 40 V range of the backgate voltage with a 1:1 relationship between $V_{\mathrm{write}}$ and the shift in the CNP (Inset, 2b). This range corresponds to induced electron densities in the graphene channel of up to $3\mathrm{x}10^{12}$ $\mathrm{cm}^2$ as determined by estimates from the device geometric capacitance.

This annealing and doping provide control over the device's quality, as noted by increased Dirac point resistance and increased sharpness with increasing $|V_{\mathrm{write}}|$ while charging (Fig. 2a). In order to quantify these changes, we employ two modeling approaches. First, we adopt the commonly used diffusive transport model from Kim *et al.*[9]

$$R = R_{\mathrm{contact}} + \frac{L}{We\mu\sqrt{n_0^2 + n^2}}$$

Here, L =12 μm and W = 5 μm are the length and width of our conductive channel, $\mu$ is the carrier mobility, $n_0$ is the residual carrier density, and $n$ is the backgate-induced carrier density, which we calculate from the capacitance between the graphene channel and Si backgate.[9,40] $R_{\mathrm{contact}}$ is extracted from the resistance at large backgate voltages (relative to the CNP) and varies between 10 Ω -100 Ω depending on the extent of the doping, being lower when the device is more heavily doped. By using the above equation, we extract the carrier mobility and residual carrier density across the range of photodoping $V_{\mathrm{write}}$ values (Fig. 2c, 2d). Upon photoannealing the device, we see an immediate sharpening in the transfer curve. By using our photodoping technique, we are then able to tune the mobility between 200,000 and 300,000 $cm^2/V$-s as $|V_{\mathrm{write}}|$ is increased. As we discharge the device, increasing $V_{\mathrm{write}}$ from -38 V to +8 V, we see the mobility approximately return to its initial value of 200,000 at $V_{\mathrm{write}} = -4\,V$ (closest to the initial value of the CNP before doping), then become suppressed at $V_{\mathrm{write}} \geq 0$.

More interestingly, this doping technique enables control over the residual carrier density in the system (Fig. 2d). This metric describes the density of Coulomb scatterers in or near the graphene channel that affect the quality of the device. Photoannealing the initial, undoped, device causes this factor to immediately decrease by half, explaining the increased sharpness of the transfer curve. This suggests that photoannealing is a meaningful approach to improve device quality, even in systems where strong doping is not desired such as Wigner crystals or other fragile correlated states that are sensitive to carrier density.[41,42] The residual carrier density remains around $10^{10}$ $cm^{-2}$ as the device is doped with increasing magnitude to $V_{\mathrm{write}}$ = -38 V. We can then reverse the effect of the procedure: the residual density remains constant as the photodoping backgate voltage is increased in steps towards $V_{\mathrm{write}}$ = -10 V, corresponding to a decrease in the n-doping of the channel. The residual density then returns to the initial value of the device (before photoannealing) at $V_{\mathrm{write}} = 4\,V$ and becomes greater than the initial value when we attempt to p-dope the channel, reaching ~$10^{11}$ $cm^{-2}$.

We propose a mechanism of correlated charged impurities trapped at the hBN/$SiO_2$ interface to explain the reduction in charge disorder. In this picture, the photodoped carriers fill trap states in a correlated manner; due to Coulombic repulsion the charges will tend to distribute uniformly. This smooths out the electron-hole puddles present in the initial device, replacing that uneven Coulomb landscape with a more uniform environment conducive to carrier transport.[40,43] As we discharge the device, the charges are de-trapped randomly, giving rise to increased disorder.[43] Additionally, our materials system and choice of illumination may lead to inconsistent or disordered p-doping of the channel. Existing studies report inconsistent results when attempting to use photodoping to achieve p-doping in similar systems, with most studies only reporting successful n-doping as we do here.[12,25,26,28,29] In studies utilizing electron beam doping, which is proposed to generate an identical form of modulation doping, p-type doping of encapsulated graphene devices was demonstrated with a similar enhancement of transport parameters and transfer curves.[32,33] This strongly suggests that difficulty inducing p-type doping and achieving transport enhancement is related to our optical excitation scheme, and utilizing short-wavelength illumination could meaningfully improve results in this area, but goes beyond the extent of our current study.

For devices of exceptional quality, the mobility in the above diffusive model may not fully capture changes in the transport behavior. In order to further understand the properties of our device we employ the Landauer transport formalism to interrogate the scattering of carriers within our tunable electrostatic landscape. The current in our device can be written as:

$$I = \frac{2e}{h}\int T(E)M(E)[f_1(E) - f_2(E)]dE$$

Which we rewrite in terms of the conductivity of graphene[44,45]:

$$G = \frac{2eW}{v_f(\hbar\pi)^2 V_{\mathrm{DS}}}\int T(E)|E|[f_1(E) - f_2(E)]dE$$

Here, $v_f$ is the Fermi velocity, $V_{\mathrm{DS}}$ is our applied source-drain voltage, $f_{1,2}(E)$ are the source and drain Fermi functions, $L$ is the length and $W$ is the width of our conductive channel. $|E|$ arises from the density of modes in graphene's bandstructure:

$$M(E) = \frac{4W|E|}{hv_f}$$

$T(E)$ is a quantity that captures the transmission probability of these modes[46]:

$$T(E) = \frac{\lambda(E)}{\lambda(E) + L}$$

with the energy dependent backscattering mean free path:

$$\lambda(E) = l_0(n)\left(\frac{E}{k_b T}\right)^s$$

Where $l_0(n)$ is a carrier density-dependent elastic scattering length and $s$ is an exponent that corresponds to the scattering mechanism. We choose $s = \frac{1}{2}$, corresponding to 2D Coulomb disorder at a distance greater than a few nm from our channel, which is consistent with trapped charges in the hBN-$SiO_2$ interface for our hBN thickness of approximately 50 nm.[47] We integrate the Landauer model numerically to extract $l_0(n)$ from our transfer curve measurements (Fig. 2e). We observe that for $V_{\text{write}} < 0$ V, corresponding to n-type doping of the channel, the scattering length at moderate to large density increases by over an order of magnitude. For transistor-like operation of such a device, this scattering dependence, which we can tune in situ through our photodoping scheme, serves to enhance the on-off ratio. We observe suppressed scattering lengths near the CNP, but the Landauer picture is insufficient to capture the minimum conductivity of graphene in this region.[48–50] Rather, we focus on quasi-ballistic transport by extracting the mean scattering length at a representative density ($n = 1.5 \times 10^{11}$ cm$^{-2}$) away from the Dirac point (Fig. 2f). We see the remarkable ability to tune the device between the quasi-ballistic and diffusive transport regimes. For our device length of ~12 μm, scattering lengths of 0.1 μm correspond to diffusive transport, while scattering lengths of 1 μm or greater, comparable to the length of the channel, correspond to quasi-ballistic transport. Thus, we have demonstrated the ability to tune the quality of our 2D material channel in situ, providing the means to achieve a quasi-ballistic test structure without additional fabrication steps.

One hallmark of reduced disorder in graphene devices is the emergence of enhanced Coulombic interactions and correlated physics. To further test the quality of our device, we sought to observe signatures of quantum Hall isospin ferromagnetic states known to form in the highest quality graphene devices.[38] We first perform quantum Hall transport measurements of the device with an applied magnetic field of 9 T in its high disorder regime, achieved after discharging the device with $V_{\text{write}} = 8$ V (Fig. 3a) for ten minutes. Measurements at high magnetic field reveal conventional quantum Hall plateaus with 4-fold degeneracy arising from the spin and valley degrees of freedom at filling factors of $\nu = -2, -6, -10$. These emerge on both the electron and hole-doped sides. We additionally plot the Landau fan diagram for the same device to fully resolve the evolution of the quantum Hall transport as a function of magnetic field, and observe no noteworthy behavior (Fig. 3b). Red lines show fitting to the Diophantine equation: $(n/n_0) = t(\phi/\phi_0) + s$. Here, $(n/n_0)$ and $(\phi/\phi_0)$ are carrier density and magnetic flux per graphene unit cell and $t, s$ are integers.[51] Even in this unideal regime, we see significant enhancement in the quantum Hall data. Before any treatment was applied to the device the 2nd and 3rd Landau levels on the hole side could not be resolved, and the resolvable minima in $R_{xx}$ were over an order of magnitude greater (Fig. S3).

We then demonstrate the ability to tune the behavior of the Landau level degeneracy in situ. We perform the photoannealing process overnight, also with $V_{\text{write}}$ = 0 V, and repeat the measurement (Fig. 3c). This process introduces no net charge but when performed for several hours allows for the rearrangement of electron-hole puddles in the Coulomb environment into a minimum disorder state. On the hole side (negative backgate voltage), we see additional plateaus emerge in $G_{xy}$ (red, left axis) and valleys in $R_{xx}$ (blue, right axis) at all nonzero integer filling factors, corresponding to full lifting of the normal 4-fold degeneracy of the Landau levels. This degeneracy lifting is caused by spontaneous symmetry breaking of the spin and valley degrees of freedom due to many-body Coulomb interactions, and its observation here indicates the exceptional quality of the device.[38,52–54] In the Landau fan diagram for this measurement (Fig. 3d), full degeneracy lifting can be observed at filling factors of $\nu = -1$ through $\nu = -14$. This degeneracy lifting is complete in the 2nd Landau level as low as 4.4 T and is completely realized in the 1st, 2nd and 3rd Landau Levels at 6.4 T, lower than in most studies that demonstrate this phenomena (where magnetic fields of 9 T or more are used), a further indication of the exceptional sample quality.[38,52,53] At 7 T, degeneracy lifting in the 0th Landau level emerges as well, and $\nu = -1$ can be resolved. Photodoping the sample from this point further enhances the electronic correlation. After photodoping, we observe degeneracy lifted states at filling factors $\nu = -1$ through $\nu = -11$ at as low as 4 T (Fig. S1). Additionally, we observe degeneracy lifting on the electron side between 3.5 T and 6 T in both the photoannealed and photodoped regimes (Fig. S1, S2). This feature is not universally observed in graphene quantum Hall devices, and usually requires very large applied fields.[38,52] Our observation of it here at low field is comparable to the highest quality graphene devices hosting these states.[54] The presence of degeneracy lifting at lower magnetic fields can be viewed as a reduction in the disorder-induced broadening, Γ, of the Landau levels. Γ corresponds to a competing energy scale arising from Coulomb disorder in the environment, analogous to the residual carrier density discussed above in the classical transport context [55]. In this case the disorder outcompetes the Zeeman or exchange energies and washes out the symmetry-breaking responsible for the quantum Hall isospin ferromagnetic states. We can roughly estimate Γ from the minimum magnetic energy at which degeneracy lifting is observed, which is on the order of 0.4 meV in our sample after photoannealing, reduced significantly from >1 meV prior (beyond our magnetic field range).

Here, we have demonstrated a means to achieve exceptional enhancement and control of a common 2D materials heterostructure. Photoannealing provides a straightforward approach to enhancing the quality of future 2D materials devices, even in studies that do not call for doping. Furthermore, photodoping enables precise control over charge in the device and its transport properties. These observations are supported by our theoretical analysis which demonstrates reversible enhancement in mobility, residual carrier density and scattering length. We establish that this control extends to in situ tuning of correlated physics via the emergence of spin and valley quantum Hall ferromagnetic states in our graphene device. These measurements indicate that our techniques provide a direct and tunable degree of control over electronic disorder in 2D devices, manifesting as significantly reduced Coulomb scattering in the device channel. Reducing sources of disorder has been key in advancing studies of sensitive physics in nanoscale systems. Reducing local electronic disorder can improve key parameters of existing technologies by multiple orders of magnitude.[56] Our photoinduced modulation doping approach addresses the problem by utilizing the naturally low *intrinsic* disorder introduced by modulation doping while simultaneously suppressing *extrinsic* disorder.[57] This enables a less stochastic and more deterministic approach to 2D materials device fabrication, where subtle but highly impactful differences in device construction can be easily homogenized. Photodoping can be applied in a wide variety of ways, for example, to solve disorder-induced variability in threshold voltages of 2D material transistors.[58] It could also be extended to Moiré heterostructures and used to stabilize fragile correlated states.[59] We also wish to highlight that this effect may have been overlooked or underreported in past, similar studies. Because photodoping effects can be accessed with broadband illumination at moderate intensities very similar to standard laboratory light sources (i.e., probe station lamps), many researchers may have inadvertently utilized photodoped devices to the benefit or detriment of their work. Our methodology provides a route to utilize photodoping and photoannealing consistently to reduce variability, realize sensitive physics, and accelerate the field of 2D materials device research.

# METHODS

## Device Fabrication

A dry transfer method was used to fabricate and transfer an exfoliated $h$BN/Gr/$h$BN heterostructure onto a p+Si/SiO2 (280 nm) substrate with pre-patterned alignment markers.[60] The heterostructure was then vacuum (~ $10^{-6}$ torr) annealed at 350 ˚C for 3 h (temperature ramp rate of 0.5 ˚C/s) to clean off any residual polymer. The graphene was a single layer and the thickness of both the top and bottom $h$BN layers was ~50 nm. The heterostructure was then patterned into a conventional Hall bar geometry by using electron-beam (e-beam) lithography followed by subsequent reactive ion etching with $CHF_3$ and $O_2$. A region free of interfacial contamination (as measured by atomic force microscopy was selected as the central region of the Hall bar, to achieve the highest device quality. A second e-beam lithography step, followed by the e-beam deposition of Cr/Pd/Au (5 nm/10 nm/75 nm) and subsequent metal lift-off, was used to form 1D metal edge contacts to the $h$BN-sandwiched graphene (Fig. 1a, c).[61]

## Transport Measurements

All photodoping and subsequent transport measurements were performed at 1.6 K in a closed-cycle magneto-optic cryostat system at NIST. We note that doping at room temperature was equally successful. For magnetotransport measurements, out-of-plane magnetic fields between 0 and 9 T were applied via a superconducting magnet. The device was only illuminated during the doping process: during measurement, all illumination sources were turned off and care was taken to keep the device in the dark. The hall bar geometry allows simultaneous measurement of $R_{xx}$ ($G_{xx}$) and $R_{xy}$ ($G_{xy}$), with currents ranging from 5 nA to 100 nA. For transfer curve measurements, the device was connected in series with a 2 MΩ resistor and a 10 kΩ resistor which act a constant current source, and the voltage drop over the device ($R_{xx}$ and $R_{xy}$) were measured as a function of the backgate voltage using a lock-in amplifier at 19 Hz.

Another lock-in unit was used to measure the voltage drop over the 10 kΩ resistor to monitor the source-drain current. Carrier densities were estimated using a standard geometric capacitance model. The contribution of the quantum capacitance was multiple orders of magnitude smaller than the geometric capacitance and was therefore ignored.

**Disclaimer**

Certain commercial equipment, instruments, software, or materials are identified in this paper in order to specify the experimental procedure adequately. These identifications are not intended to imply recommendation or endorsement by the National Institute of Standards and Technology (NIST), nor are they intended to imply that the materials or equipment identified are necessarily the best available for the purpose.

This document has not been peer reviewed but has been cleared by NIST for release.

**Acknowledgments**

The authors from LPS gratefully acknowledge assistance from the LPS support staff including G Latini, J Wood, R Brun, P Davis and D Crouse. MFM and RT would like to acknowledge the NIST/National Research Council Postdoctoral Research Associateship Program for funding. This work was performed with funding from the CHIPS Metrology Program, part of CHIPS for America, National Institute of Standards and Technology, U.S. Department of Commerce.

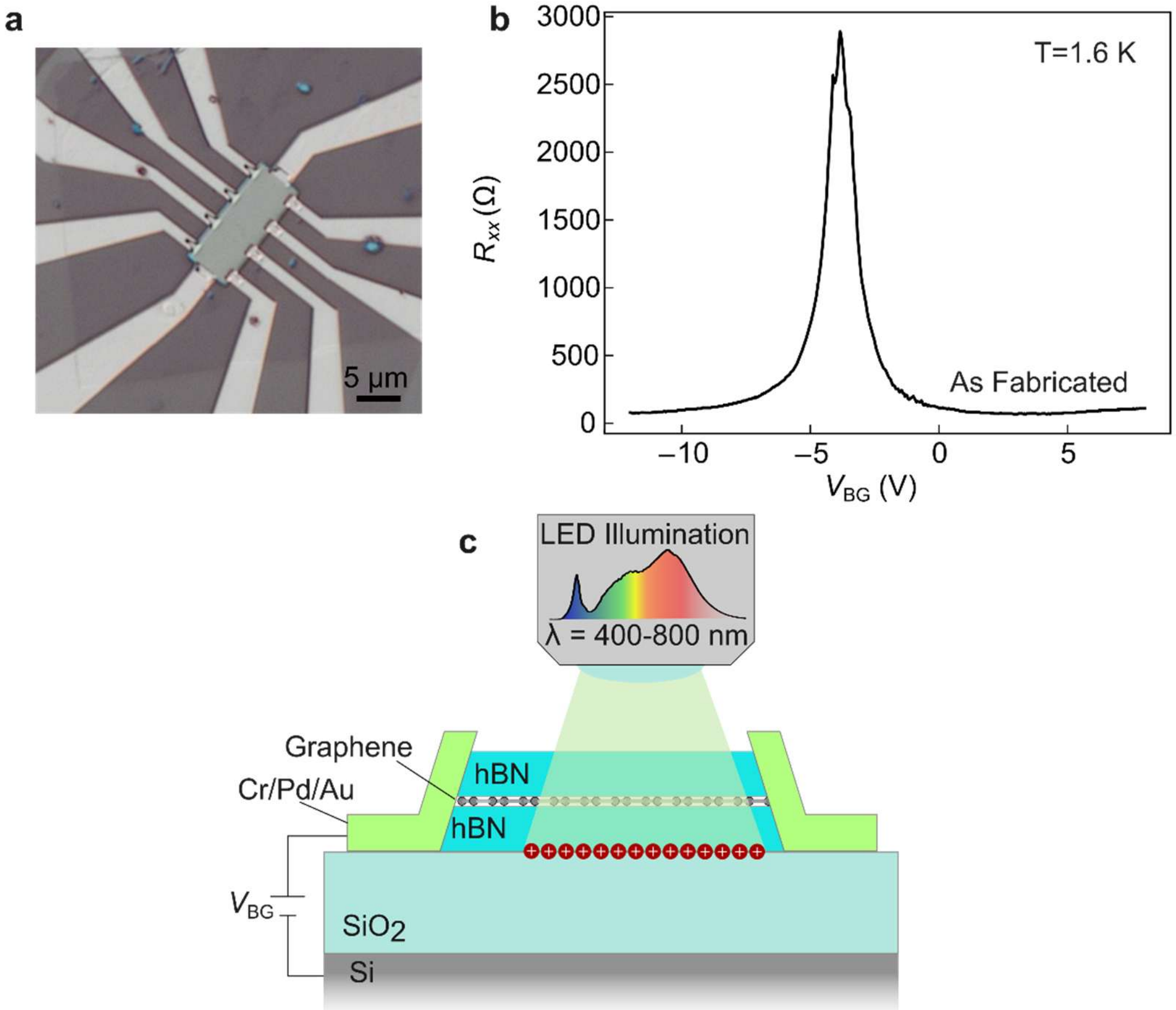

Figure 1: (a) Optical Image of encapsulated graphene device used in this work, where heavily doped Si can be used as a global backgate. Devices were fabricated with standard dry transfer and electron beam lithography techniques. (b) Initial transfer curve measurement (T = 1.6 K) of the device, before any photodoping is applied. (c) Device and photodoping schematic. Under illumination, photoexcited carriers are trapped at the hBN-SiO2 interface, leading to long-lived modulation doping of the graphene channel.

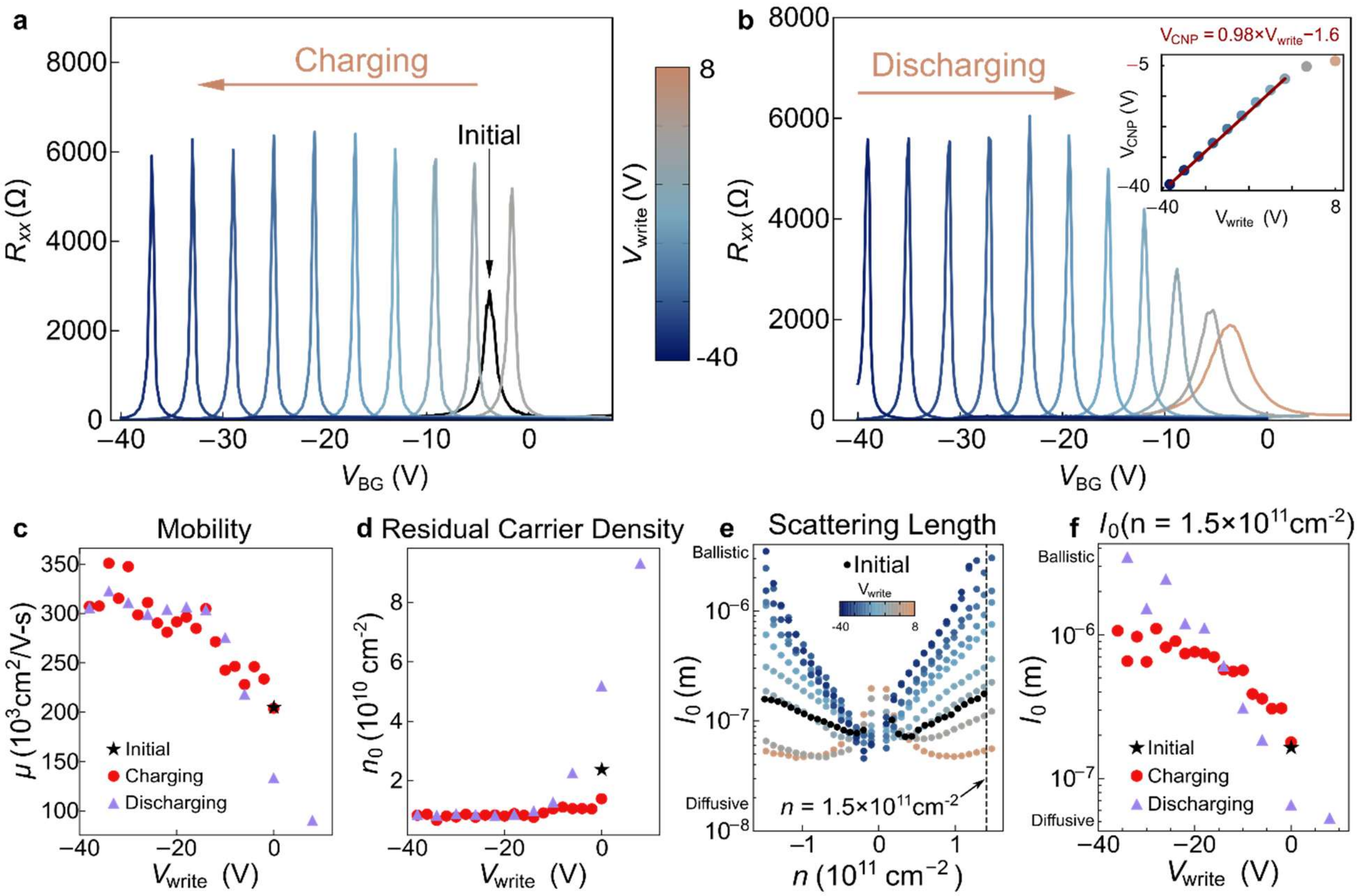

Figure 2: (a) Transfer curve measurements of encapsulated graphene device on SiO2, with increasing magnitude of photodoping voltage $V_{write}$. (b) Measurements of the same device with descending photodoping voltage, including one measurement of positive $V_{write}$ = 8 V, attempting to achieve p-type doping. Inset: CNP of the device can be controlled one-to-one with negative values of $V_{write}$. (c,d) Extracted mobility and residual carrier density from the curves in a and b, showing device enhancement upon photoannealing and photodoping. When discharging, the device can be brought back to its initial configuration and further doping degrades the mobility and continues to increase the residual carrier density beyond the initial value. The initial (pre-photoannealing) data point is placed at $V_{write}$ = 0 V for comparison. (e) Extracted scattering lengths as a function of carrier density from our modified Landauer transport model. Dashed line shows carrier density highlighted in f. (f) Scattering lengths away from the Dirac point at a representative carrier density of 1.5x10$^{11}$ cm$^{2}$. This scattering length can be reversibly tuned from the diffusive to quasi-ballistic regime via our photodoping method.

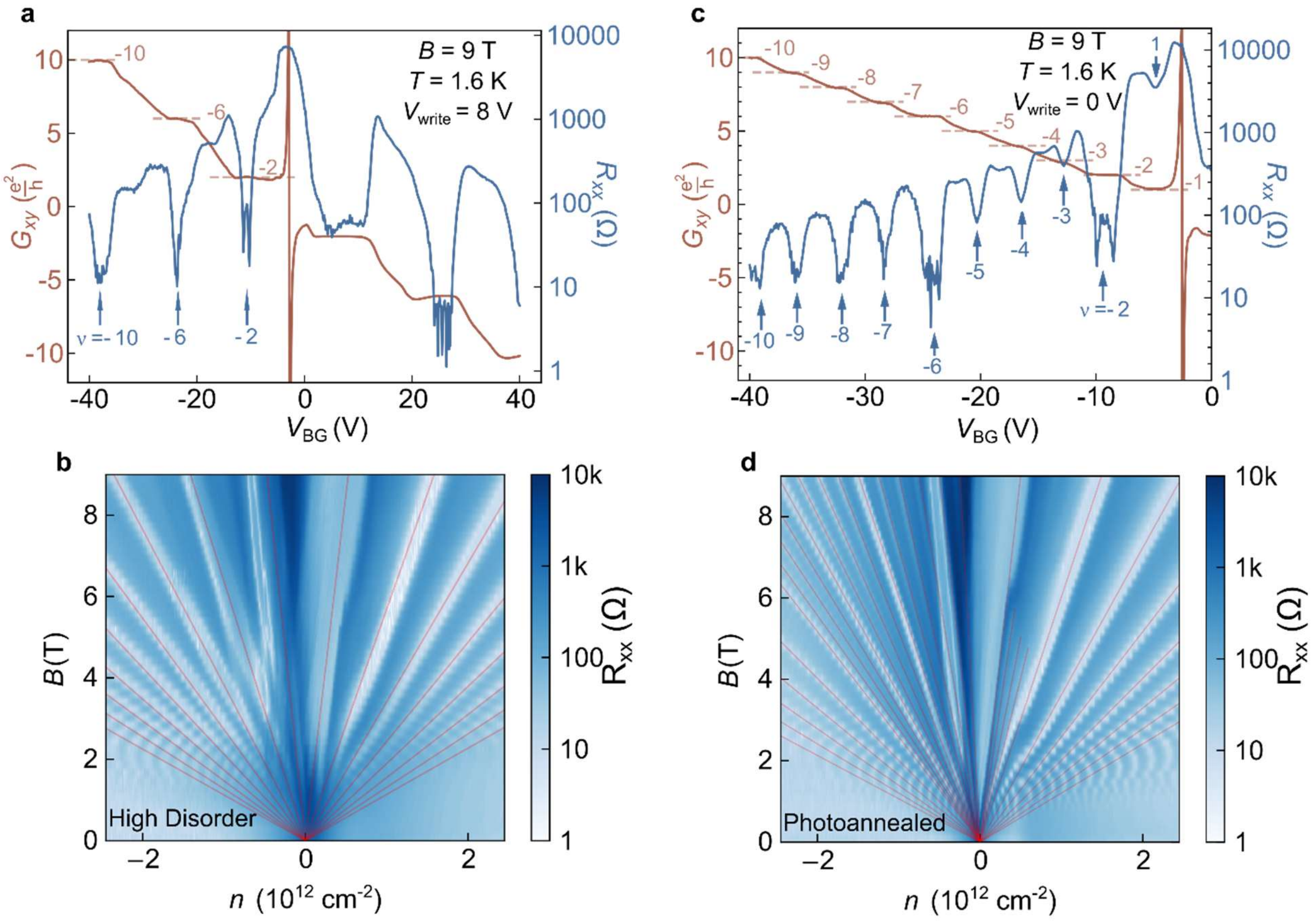

Figure 3: Quantum Hall transport measurements before and after photoannealing the graphene device. (a) Hall conductivity (Red – left axis) and longitudinal resistance (Blue – right axis) for the device in the high disorder regime at 9 T and 1.6 K. These measurements show expected 4-fold degenerate plateaus in $G_{xy}$ at filling factors of 2,6, and 10. Minima in $R_{xx}$ are labelled with the corresponding filling factor in blue. (b) Landau fan diagram for the measurement in a, showing the evolution of $R_{xx}$ as a function of magnetic field. (c) The same Quantum Hall measurements for the device after it is photoannealed. Both Hall conductivity and longitudinal resistance demonstrate complete degeneracy lifting on the hole side (negative backgate voltage), where the normally 4-fold degenerate landau levels split into 4 distinct plateaus. This behavior is additionally seen in the 1st Landau level on the electron side (Figure S2). (d) Landau fan diagram for the measurement in b. Degeneracy lifting can be seen as low as 3.5 T, with complete lifting in the first 3 Landau levels on the hole side at 6.4 T and greater. Splitting in the 0th Landau level can be seen at 7 T and greater, where the $\nu = 1$ plateau emerges. Splitting of the 1st Landau level in the electron side can be seen at 3.5 T, a further indication of exceptionally low disorder.

# Supplemental Information

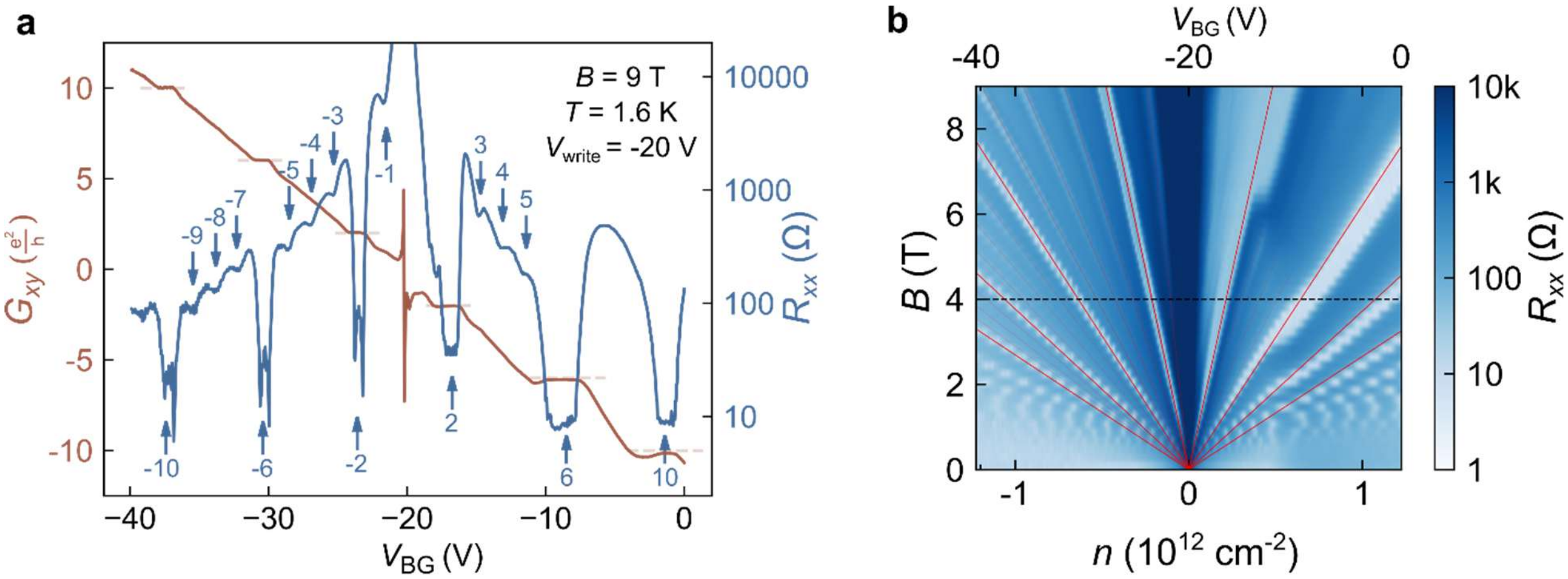

Figure S1: Quantum Hall measurements in the low Coulomb disorder regime. The device is photodoped with $V_{write}$ = -20V. (a) Full degeneracy lifting in the 0th, 1st and 2nd Landau levels on the hole side can be observed in $R_{xx}$ at magnetic fields as low as 4T, indicating further enhancements in quality relative to the photoannealed sample in Figure 3. Full degeneracy lifting is also observed in the 1st Landau level on the electron side at 4T. (b) Landau fan diagram for the highly photodoped device, with fitting to the Diophantine equation in red. Dashed line at 4T shows the location of the linecut in a.

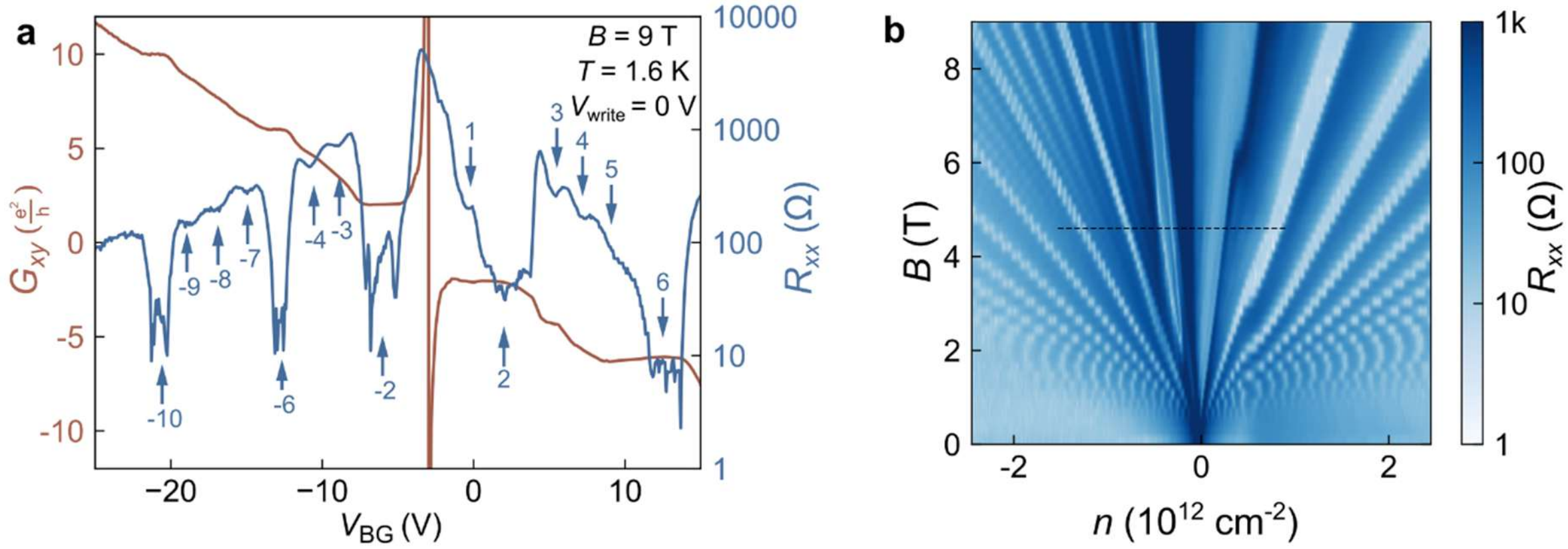

Figure S2: Quantum Hall ferromagnetic states of electrons and holes in the photoannealed device. (a) Extended plot of figure 3B, with the electron-doped side shown. States at $\nu = 1, 3$, and $4$ are clearly resolved in $R_{xx}$ at 4.6 T, indicative of exceptional sample quality and low local disorder. (b) Landau fan diagram for the photoannealed device, with black dashed line indicating the linecut shown in b. The symmetry broken state at $\nu = 5$ can be observed in a small range of magnetic fields below the highlighted region. The color scale is changed and fittings are removed to highlight the features on the electron side.

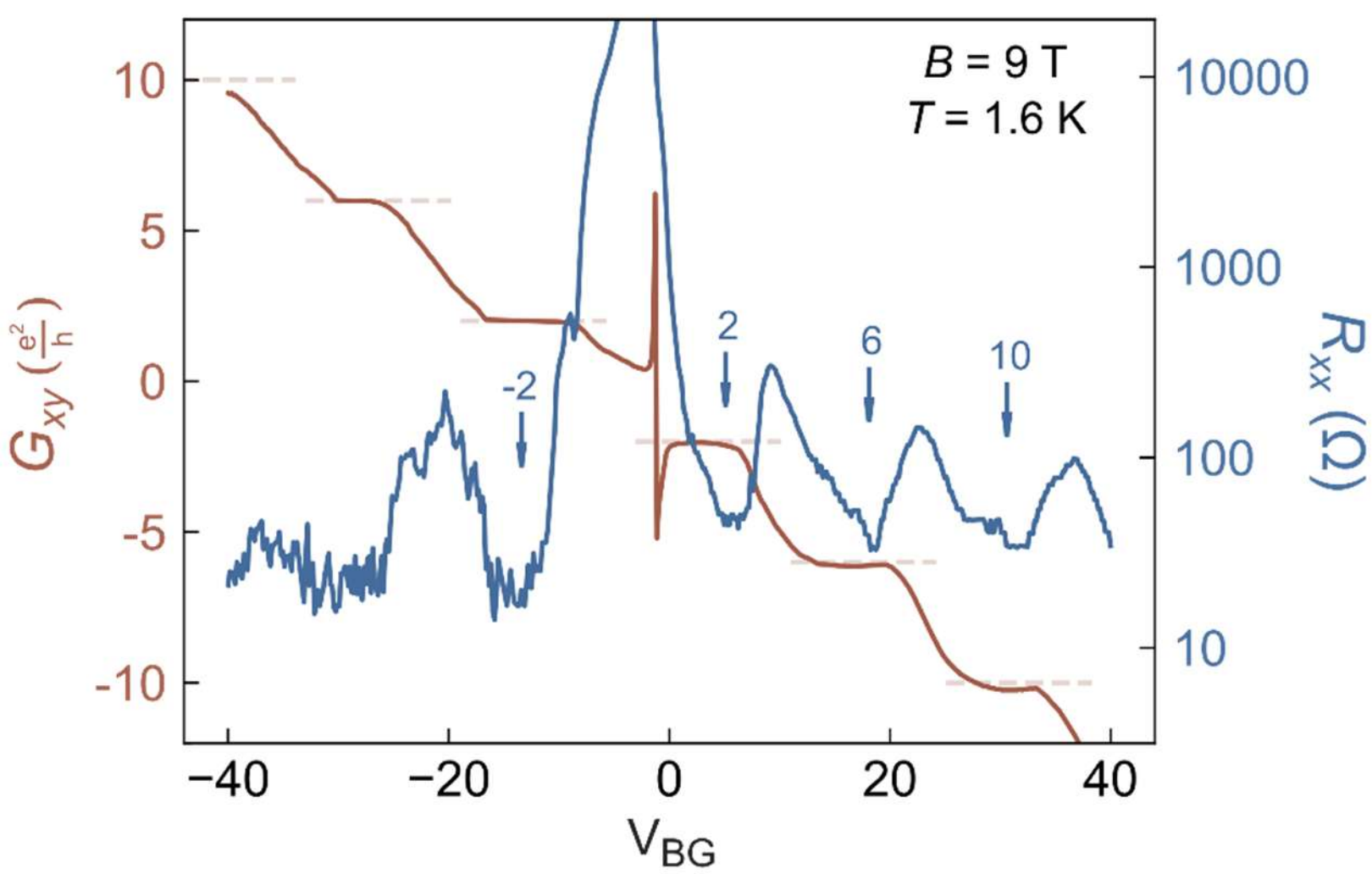

Figure S3. Quantum Hall measurements of the device before initial photoannealing treatment. Device shows broad 4-fold degenerate plateaus in $G_{xy}$ on both the electron and hole sides. $R_{xx}$ data appears highly noisy with minima around 40 Ω at only $\nu = -2, 2, 6, 10$.

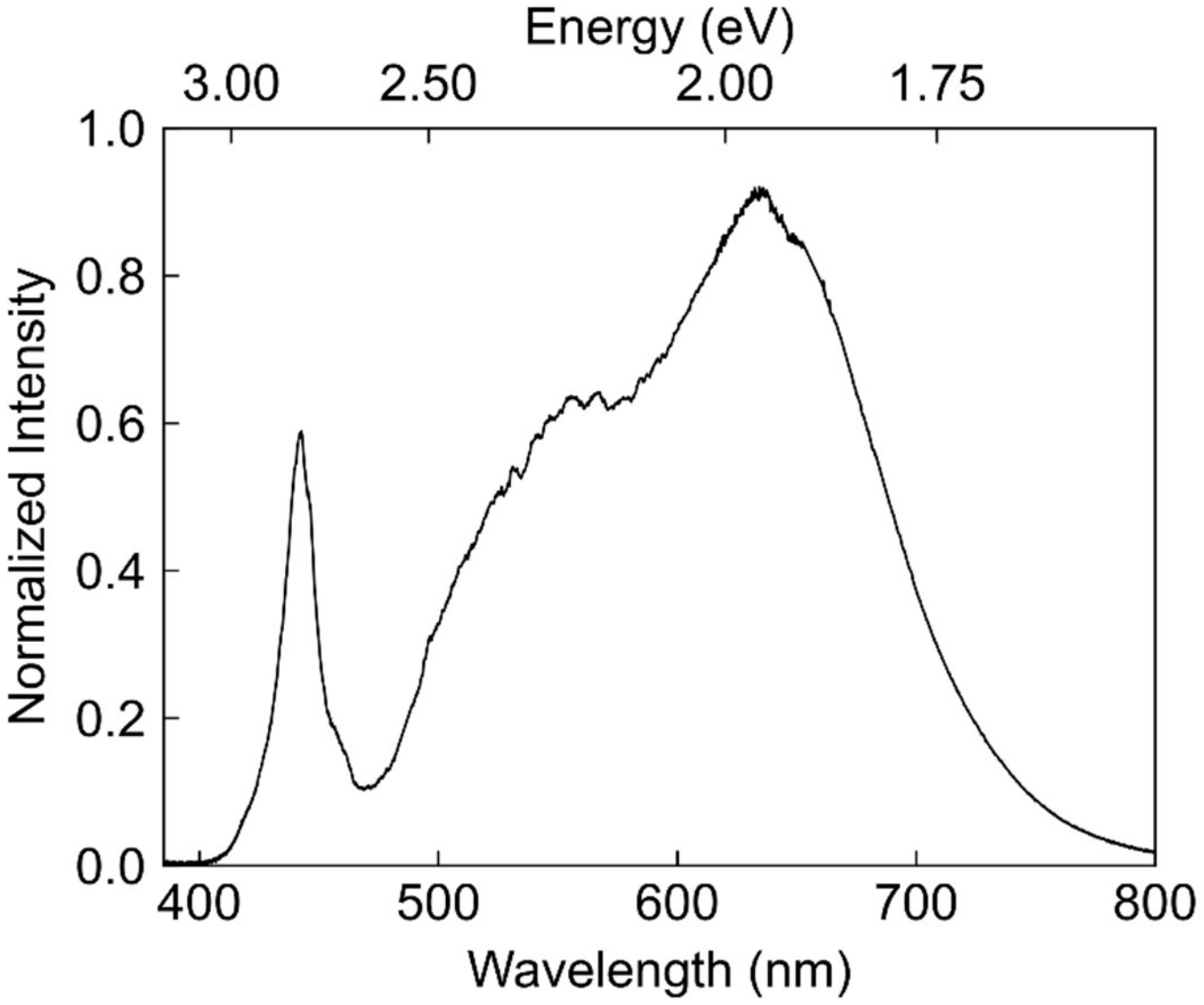

Figure S4. Emission Spectrum of the Thorlabs MWWHL3 light source used in this work. Data was provided by Thorlabs.

1. Ullah, S. *et al.* Advances and Trends in Chemically Doped Graphene. *Adv. Mater. Interfaces* **7**, 2000999 (2020).
2. Liu, H., Liu, Y. & Zhu, D. Chemical doping of graphene. *J. Mater. Chem.* **21**, 3335–3345 (2011).
3. Schedin, F. *et al.* Detection of individual gas molecules adsorbed on graphene. *Nat. Mater.* **6**, 652–655 (2007).
4. Jang, J. *et al.* Reduced dopant-induced scattering in remote charge-transfer-doped MoS2 field-effect transistors. *Sci. Adv.* **8**, eabn3181 (2022).
5. Lin, P.-C. *et al.* Doping Graphene with Substitutional Mn. *ACS Nano* **15**, 5449–5458 (2021).
6. Joshi, P., Romero, H. E., Neal, A. T., Toutam, V. K. & Tadigadapa, S. A. Intrinsic doping and gate hysteresis in graphene field effect devices fabricated on $SiO_2$ substrates. *J. Phys. Condens. Matter* **22**, 334214 (2010).
7. Novoselov, K. S. *et al.* Electric Field Effect in Atomically Thin Carbon Films. *Science* **306**, 666–669 (2004).
8. Morozov, S. V. *et al.* Giant Intrinsic Carrier Mobilities in Graphene and Its Bilayer. *Phys. Rev. Lett.* **100**, 016602 (2008).
9. Kim, S. *et al.* Realization of a high mobility dual-gated graphene field-effect transistor with Al2O3 dielectric. *Appl. Phys. Lett.* **94**, 062107 (2009).
10. Guo, B., Fang, L., Zhang, B. & Gong, J. R. Graphene Doping: A Review. *Insciences J* **1**, 80–89 (2011).
11. Neilson, K. *et al.* Threshold Voltage Control through Solvent Doping of Monolayer MoS2 Transistors. *Nano Lett.* **25**, 7778–7784 (2025).
12. Ju, L. *et al.* Photoinduced doping in heterostructures of graphene and boron nitride. *Nat. Nanotechnol.* **9**, 348–352 (2014).

13. Le, S. T. *et al.* Spatial photoinduced doping of graphene/hBN heterostructures characterized by quantum Hall transport. *2D Mater.* **12**, 015006 (2025).

14. Zhang, Z. *et al.* Photodoping strategies in two-dimensional semiconductors: Mechanisms, characterizations, and emerging applications. *InfoMat* **n/a**, e70092 (2025).

15. Epping, A. *et al.* Quantum transport through $MoS_2$ constrictions defined by photodoping. *J. Phys. Condens. Matter* **30**, 205001 (2018).

16. Gadelha, A. C. *et al.* Local photodoping in monolayer MoS2. *Nanotechnology* **31**, 255701 (2020).

17. Dingle, R., Störmer, H. L., Gossard, A. C. & Wiegmann, W. Electron mobilities in modulation-doped semiconductor heterojunction superlattices. *Appl. Phys. Lett.* **33**, 665–667 (1978).

18. Pfeiffer, L., West, K. W., Stormer, H. L. & Baldwin, K. W. Electron mobilities exceeding $10^7$ $cm^2$/V s in modulation-doped GaAs. *Appl. Phys. Lett.* **55**, 1888–1890 (1989).

19. Biswas, S., Li, Y., Winter, S. M., Knolle, J. & Valentí, R. Electronic Properties of RuCl3 in Proximity to Graphene. *Phys. Rev. Lett.* **123**, 237201 (2019).

20. Wang, Y. *et al.* Modulation Doping via a Two-Dimensional Atomic Crystalline Acceptor. *Nano Lett.* **20**, 8446–8452 (2020).

21. Xie, J. *et al.* Low Resistance Contact to P-Type Monolayer WSe2. *Nano Lett.* **24**, 5937–5943 (2024).

22. Wang, D., Li, X.-B. & Sun, H.-B. Modulation Doping: A Strategy for 2D Materials Electronics. *Nano Lett.* **21**, 6298–6303 (2021).

23. Lee, D. *et al.* Remote modulation doping in van der Waals heterostructure transistors. *Nat. Electron.* **4**, 664–670 (2021).

24. Neumann, C. *et al.* Spatial Control of Laser-Induced Doping Profiles in Graphene on Hexagonal Boron Nitride. *ACS Appl. Mater. Interfaces* **8**, 9377–9383 (2016).

25. Choi, H. H. *et al.* Photoelectric Memory Effect in Graphene Heterostructure Field-Effect Transistors Based on Dual Dielectrics. *ACS Photonics* **5**, 329–336 (2018).

26. Wu, E. *et al.* Dynamically controllable polarity modulation of MoTe2 field-effect transistors through ultraviolet light and electrostatic activation. *Sci. Adv.* **5**, eaav3430 (2019).

27. Li, S. *et al.* A High-Performance In-Memory Photodetector Realized by Charge Storage in a van der Waals MISFET. *Adv. Mater.* **34**, 2107734 (2022).

28. Wang, S. *et al.* Nonvolatile van der Waals Heterostructure Phototransistor for Encrypted Optoelectronic Logic Circuit. *ACS Nano* **16**, 4528–4535 (2022).

29. Xiang, D. *et al.* Two-dimensional multibit optoelectronic memory with broadband spectrum distinction. *Nat. Commun.* **9**, 2966 (2018).

30. Pan, X. *et al.* Parallel perception of visual motion using light-tunable memory matrix. *Sci. Adv.* **9**, eadi4083 (2023).

31. Sczygelski, E. *et al.* Extrinsic and intrinsic photoresponse in monodisperse carbon nanotube thin film transistors. *Appl. Phys. Lett.* **102**, 083104 (2013).

32. Lan, Q. *et al.* Rewritable Complementary Nanoelectronics Enabled by Electron-Beam Programmable Ambipolar Doping. Preprint at https://doi.org/10.48550/arXiv.2512.06318 (2025).

33. Shi, W. *et al.* Reversible writing of high-mobility and high-carrier-density doping patterns in two-dimensional van der Waals heterostructures. *Nat. Electron.* **3**, 99–105 (2020).

34. Li, S. *et al.* Light-Rewritable Logic Devices Based on Van der Waals Heterostructures. *Adv. Electron. Mater.* **8**, 2100708 (2022).

35. Liu, T. *et al.* Nonvolatile and Programmable Photodoping in MoTe2 for Photoresist-Free Complementary Electronic Devices. *Adv. Mater.* **30**, 1804470 (2018).

36. Aftab, S., Akhtar, I., Seo, Y. & Eom, J. WSe2 Homojunction p–n Diode Formed by Photoinduced Activation of Mid-Gap Defect States in Boron Nitride. *ACS Appl. Mater. Interfaces* **12**, 42007–42015 (2020).

37. Tsai, M.-Y. *et al.* A reconfigurable transistor and memory based on a two-dimensional heterostructure and photoinduced trapping. *Nat. Electron.* **6**, 755–764 (2023).

38. Young, A. F. *et al.* Spin and valley quantum Hall ferromagnetism in graphene. *Nat. Phys.* **8**, 550–556 (2012).

39. Quereda, J., Ghiasi, T. S., van der Wal, C. H. & van Wees, B. J. Semiconductor channel-mediated photodoping in h-BN encapsulated monolayer MoSe2 phototransistors. *2D Mater.* **6**, 025040 (2019).

40. Adam, S., Hwang, E. H., Galitski, V. M. & Das Sarma, S. A self-consistent theory for graphene transport. *Proc. Natl. Acad. Sci.* **104**, 18392–18397 (2007).

41. Xu, Y. *et al.* Correlated insulating states at fractional fillings of moiré superlattices. *Nature* **587**, 214–218 (2020).

42. Gupta, S., Kutana, A. & Yakobson, B. I. Heterobilayers of 2D materials as a platform for excitonic superfluidity. *Nat. Commun.* **11**, 2989 (2020).

43. Yan, J. & Fuhrer, M. S. Correlated Charged Impurity Scattering in Graphene. *Phys. Rev. Lett.* **107**, 206601 (2011).

44. Sanchez Esqueda, I. & Cress, C. D. Modeling Radiation-Induced Scattering in Graphene. *IEEE Trans. Nucl. Sci.* **62**, 2906–2911 (2015).

45. Esqueda, I. S. *et al.* The impact of defect scattering on the quasi-ballistic transport of nanoscale conductors. *J. Appl. Phys.* **117**, 084319 (2015).

46. Lundstrom, M. & Jeong, C. *Near-Equilibrium Transport: Fundamentals and Applications*. (World Scientific, 2012).

47. Das Sarma, S. & Hwang, E. H. Universal density scaling of disorder-limited low-temperature conductivity in high-mobility two-dimensional systems. *Phys. Rev. B* **88**, 035439 (2013).

48. Khveshchenko, D. V. Electron Localization Properties in Graphene. *Phys. Rev. Lett.* **97**, 036802 (2006).

49. Nomura, K. & MacDonald, A. H. Quantum Transport of Massless Dirac Fermions. *Phys. Rev. Lett.* **98**, 076602 (2007).

50. Sui, Y., Low, T., Lundstrom, M. & Appenzeller, J. Signatures of Disorder in the Minimum Conductivity of Graphene. *Nano Lett.* **11**, 1319–1322 (2011).

51. Dean, C. R. *et al.* Hofstadter's butterfly and the fractal quantum Hall effect in moiré superlattices. *Nature* **497**, 598–602 (2013).

52. Zhang, Y. *et al.* Landau-Level Splitting in Graphene in High Magnetic Fields. *Phys. Rev. Lett.* **96**, 136806 (2006).

53. Chiappini, F. *et al.* Lifting of the Landau level degeneracy in graphene devices in a tilted magnetic field. *Phys. Rev. B* **92**, 201412 (2015).

54. Zimmermann, K. *et al.* Tunable transmission of quantum Hall edge channels with full degeneracy lifting in split-gated graphene devices. *Nat. Commun.* **8**, 14983 (2017).

55. Yang, C. H., Peeters, F. M. & Xu, W. Landau-level broadening due to electron-impurity interaction in graphene in strong magnetic fields. *Phys. Rev. B* **82**, 075401 (2010).

56. Domaretskiy, D. *et al.* Proximity screening greatly enhances electronic quality of graphene. *Nature* **644**, 646–651 (2025).

57. Rhodes, D., Chae, S. H., Ribeiro-Palau, R. & Hone, J. Disorder in van der Waals heterostructures of 2D materials. *Nat. Mater.* **18**, 541–549 (2019).

58. Lanza, M., Smets, Q., Huyghebaert, C. & Li, L.-J. Yield, variability, reliability, and stability of two-dimensional materials based solid-state electronic devices. *Nat. Commun.* **11**, 5689 (2020).

59. Nuckolls, K. P. & Yazdani, A. A microscopic perspective on moiré materials. *Nat. Rev. Mater.* **9**, 460–480 (2024).

60. Le, S. T. *et al.* Assembly of High-Performance van der Waals Devices Using Commercial Polyvinyl Chloride Films. Preprint at https://doi.org/10.48550/arXiv.2505.08579 (2025).

61. Wang, L. *et al.* One-Dimensional Electrical Contact to a Two-Dimensional Material. *Science* **342**, 614–617 (2013).